\documentclass[10pt]{article}
\usepackage{geometry}                
\usepackage[parfill]{parskip}
\usepackage{amssymb}

\usepackage{amsmath}

\title{Parametric evolution, addition of bound states and generalized Lax hierarchies}
\author{C. V. Sukumar \\{\em The Rudolf Peierls Centre for Theoretical Physics,}\\{\em Department of Physics, University of Oxford, Oxford OX1 3NP, U.K. }}

\begin{document}
\maketitle

\begin{abstract}
The connection of the 'time' evolution of the eigenstates of the reflectionless potentials of the Lax hierarchy to the more general case of the 'time' evolution of the eigenstates of the Schr{\"{o}}dinger equation for potentials with non-vanishing reflection coefficients is explored. A new hierarchy of functions satisfying 'time' dependent equations is established.

\end{abstract}

\section{Introduction}

The Korteweg-deVries (KdV) equation was first discovered in the study of water waves. The KdV and the related equations with higher order nonlinearity, which are members of the Lax hierarchy (Lax 1968), have played a fundamental role in the study of nonlinear systems because they simulate many physical systems (Scott {\it et al} 1973), admit many conservation laws and the multi-soliton solutions can be given in analytic form. The KdV equation and its generalisation Kadomtsev-Petviashvli (KP) equation have also played an important role in pure mathematics because of their connection to algebraic curves, Jacobian varieties, vector bundles on curves, Schur polynomials and infinite dimensional Grassmannians (Mulase 1984). 

The connection between N-soliton solutions of the KdV equation 
\begin{equation}
\frac{\partial U}{\partial t_1}\ +\ \frac{\partial^3 U}{\partial x^3}\ -\ 6 U\ \frac{\partial U}{\partial x} \ =\ 0 \ .\label{}
\end{equation}
and reflectionless potentials with $N$ bound states in non-relativistic Quantum Mechanics is well known (Kay and Moses 1956, Gardner {\it et al} 1967, Scott {\it et al} 1973 ,Thacker {\it et al} 1978). The KdV equation is the $m=1$ member of the Lax hierarchy of equations (Lax 1968, Caudrey {\it et al} 1976) defined by
\begin{equation}
\frac{\partial U}{\partial t_m}\ + \frac{\partial L_m}{\partial x}\ =\ 0 \label{}
\end{equation}
where $[L_j]$ satisfy
\begin{equation}
L_0\ =\ U\ ,\ 
\Big(\frac{\partial^3}{\partial x^3} \ - 4 U \ \frac{\partial }{\partial x}\ -\ 2 \frac{\partial U}{\partial x}\Big)\ L_{j-1}\ =\ \frac{\partial L_j}{\partial x}\ ,\ \ j=1,2,..,m \label{}
\end{equation}
and $t_m$ is the 'time' parameter of the $m^{th}$ member of the hierarchy.
The solutions to the non-linear equations of the Lax hierarchy may be used as potentials in linear Schr{\"{o}}dinger equations and their spectral properties may be studied by solving (using units in which mass = 1/2 and $\hbar =1$)
\begin{equation}
H\ =\ -\frac{\partial^2}{\partial x^2}\ +U \ \ ,\ H\ \psi_k \ =\ E_k\ \psi_k  \ . \label{}
\end{equation} 
The N-soliton solution of eq. (2) may be viewed as a reflectionless potential $U$ which supports $N$ bound states of the Hamiltonian operator $H$. Lax (1968) has shown that when the 'time' evolution of U is governed by eqs. (2) and (3), and the 'time' evolution of the eigenstates $\psi_k$ is governed  by
\begin{equation}
B_m \ \psi_k \ =\ i\frac{\partial \psi_k}{\partial t_m} \ \ ,\ \ \psi_k(x,t)\ =\ \exp(-iB_mt)\ \psi_k(x,0)\ ,\label{}
\end{equation}
where the operator $B_m$ satisfies the commutator relation
\begin{equation}
[B_m,H]\ =\ -i\frac{\partial L_m}{\partial x}\ =\ i\frac{\partial U}{\partial t_m}\ =\  i\frac{\partial H}{\partial t_m}  \ ,\label{}
\end{equation}
then the eigenvalues $E_k$ of $H$ are time independent and the eigenstates remain normalized, but the normalization constants of the bound states and the reflection and transmission coefficients acquire a time dependence. For the case $m=1$, which leads to the third order KdV, the explicit form of $B_1$ is given by
\begin{equation}
B_1\ =\ i\ \Big(-4 \frac{\partial^3}{\partial x^3}\ +\ 6 U\ \frac{\partial}{\partial x} \ +\ 3 \frac{\partial U}{\partial x}\Big)\ \ . \label{}
\end{equation}
If the potential evolves in $t_1$ according to eq. (1) and the eigenstates evolve in $t_1$ according to eqs. (5) and (7) then the eigenvalues $[E_k=-\gamma_k^2]$ in eq. (4) are independent of $t_1$, but the normalization constants $[C_k]$ which determine the behaviour of $[\psi_k]$ as $x\to\pm\infty$ and the reflection coefficient $R$ for positive energies $E=k^2$  depend on $t_1$ as given by (Scott {\it et al} 1973)
\begin{equation}
C_k(\gamma_k,t_1)\ =\ C_k(\gamma_k,0)\ \exp{\big(-4\ \gamma_k^3\ t_1)}\ ,\ \ R(k,t_1)\ =\ R(k,0)\ \exp{\big(8\ i\ k^3\ t_1\big)}\ .\label{}
\end{equation}
Similar results hold for other members of the Lax hierarchy with appropriate energy dependent changes to $C_k(\gamma_k,t_m)$ and $R(k,t_m)$.

$B_m$ is a Hermitian operator which may be interpreted as the generator of 'time' evolution which propagates the potential $U$ according to the KdV equation or another higher order non-linear equation arising from eqs. (2) and (3). This propagation of $U$ is distinct from the usual time evolution of the Schr{\"{o}}dinger eigenstates by the Hamiltonian H which propagates the particle through a fixed potential. The hermiticity of $B_m$ ensures unitary 'time' evolution of the eigenstates of the Schr{\"{o}}dinger equation.
  
In this paper we develop methods by which is is possible to establish some general results which are valid in general and the Lax hierarchy is a special case of these general results. It will be shown that it is possible to construct a hierarchy of inter-connected functions which satisfy an evolution equation in parameter space which is different from eq. (2) but reduces eq. (2) for the case of the reflectionless potentials in a suitably taken limit. 

In section 2 of the paper we outline the steps involved in a known procedure for starting from a potential which supports no eigenstates but has non-vanishing reflection coefficient to construct a potential with bound states at specified energies and scattering properties related to those of the original potential in a well defined manner. In section 3 we study how a parametric evolution different from eqs. (2) and (3) may be realized and establish some general equations. In section 4 we discuss an example of a parametric evolution equation which may be related to the evolution of a physically meaningful quantity.

When referring to 'time' in the rest of this paper we mean a parameter of the type which appears in evolution equations such as the KdV equation for the propagation of the potential. We do not refer to the time coordinate $t$ which is canonically conjugate to the energy in Quantum Mechanics and appears in the evolution equation generated by the Hamiltonian. With this clarification we drop the quotation marks when referring to 'time' in the rest of the discussion.

\section{Construction of potentials with non-vanishing reflection coefficients}


The algorithm for the construction of a reflectionless potential $U$ with $N$ bound states starting from free particle states (Thacker {\it et al 1978}) is a special case of a more general problem of starting from the solutions in a potential $U_0$ which supports no bound states and satisfies $U_0\to 0$ as $x\to\pm\infty$ and then finding a potential with bound states at specific energies (Abraham and Moses 1980, Sukumar 1986, 1987, Baye 1987, Baye and Sparenberg 1994). Let $\phi_k$ be a solution of the Schr{\"{o}}dinger equation for the potential $U_0$ at energy $E_k=-\gamma_k^2$ given by
\begin{equation}
H_0\ =\ -\frac{\partial^2}{\partial x^2}\ +  U_0\ ,\ H_0 \phi_k\ =\ \gamma_k^2\ \phi_k \label{}
\end{equation}
which goes to $0$ exponentially as $x\to\infty$ but grows exponentially as $x\to-\infty$ and therefore unnormalizable. A set of such solutions $[\phi_k]$ at energies $[-\gamma_k^2,\ k=1,2,..N]$, may be used to define a matrix $[A]$
 with elements
\begin{equation}
A_{kl} \ = \delta_{kl} \ - \ {\int_{\infty}^x \ \phi_k(y) \ \phi_l(y) \ dy }\ ,\ k,l=1,2,...N \ .\label{}
\end{equation}
The solutions $[\psi_l]$ to the system of linear equations
\begin{equation}
\sum_{l=1}^N{A_{kl}}\ \psi_l\ =\ \phi_k \ \ ,\ k=1,2,..,N
\end{equation}
may be used to define
\begin{equation}
W(x) \equiv -\sum_{l=1}^N \phi_l(x) \psi_l(x) = -\sum_{l=1}^N\sum_{k=1}^N\phi_l\big[A^{-1}\big]_{lk}\phi_k = \sum_{l=1}^N\sum_{k=1}^N\big[A^{-1}\big]_{lk} \frac{\partial A_{lk}}{\partial x} = \frac{\partial}{\partial x} {\ln{\det{A}}} .\label{}
\end{equation}
Eq. (11) may be written in the form
\begin{equation}
\psi_k(x)\ =\ \phi_k(x)\ +\ \sum_{l=1}^N \int_{\infty}^{x} \phi_k(y) \phi_l(y) \ dy\ \psi_l(x) \ ,\ \ k=1,2,...,N \label{}
\end{equation}
and can be differentiated twice and rearranged to give
\begin{align}
\big(H_0 + \gamma_k^2\big) \psi_k &= \sum_{l=1}^N \int_{\infty}^{x} \phi_k(y) \phi_l(y) dy \big(H_0 + \gamma_k^2\big) \psi_l - \sum_{l=1}^N\Big(2\phi_k \phi_l \frac{\partial \psi_l}{\partial x} + \psi_l\frac{\partial}{\partial x} (\phi_k\phi_l)\Big) \notag\\
&= \sum_{l=1}^N \int_{\infty}^{x} \phi_k(y) \phi_l(y) dy \big(H_0 + \gamma_k^2\big) \psi_l - 2\phi_k \frac{\partial}{\partial x}\sum_{l=1}^N\phi_l \psi_l - \sum_{l=1}^N\Big(\frac{\partial \phi_k}{\partial x}\phi_l - \phi_k\frac{\partial \phi_l}{\partial x}\Big) \psi_l . \label{}
\end{align}
Using the relation
\begin{equation}
\frac{\partial}{\partial x} \Big(\frac{\partial \phi_k}{\partial x} \phi_l \ -\ \phi_k \frac{\partial \phi_l}{\partial x}\Big)\ =\ \big(\gamma_k^2\ -\ \gamma_l^2\big) \ \phi_k \phi_l \label{}
\end{equation}
and eqs. (12) and (13) it can be established that
\begin{equation}
\Big(H_0\ -\ 2\frac{\partial W}{\partial x}\ +\ \gamma_k^2\Big) \psi_k(x)\ =\ \sum_{l=1}^N \int_{\infty}^{x} \phi_k(y) \phi_l(y)\ dy \Big(H_0\ -\ 2\frac{\partial W}{\partial x}\  +\  \gamma_l^2\Big) \psi_l(x) \ \label{}
\end{equation}
which may be brought to the form
\begin{equation}
\sum_{l=1}^N A_{kl}\ \big(H_0\ -\ 2\frac{\partial W}{\partial x} \ +\ \gamma_l^2\big)\ \psi_l\ =\ 0 \ .\label{}
\end{equation}
We can therefore infer that, if $\det A \ne 0$, then
\begin{align}
H_N \ &=\ H_0\ +\ U_N \ -\ U_0\\
U_N \ -U_0\ &=\ -\ 2\ \frac{\partial W}{\partial x} \ =\ -\ 2 \frac{\partial^2}{\partial x^2}{\ln{\det A}} \\
\big(H_N\ +\ \gamma_l^2\big) \psi_l \ &=\ 0 \ ,\ l=1,2,...,N \label{}
\end{align}
({\it i.e.}) $[\psi_l]$ are solutions of the Schr{\"{o}}dinger equation in the potential $U_N$ for the energy $-\gamma_l^2$. Using eqs. (10) and (11) it can be shown that $[\psi_l]$ satisfy bound state boundary conditions at $x\to\pm\infty$ establishing $[\psi_l]$ as true bound states of $U_N$.

It can also be shown that eq. (13) may be used to define $\psi(E,x)$ for positive energies $E=k^2$ by the replacements $\psi_k\to\psi(E)$ and $\phi_k\to\phi(E)$. With these replacements eq. (16) is still valid and hence, if eq. (20) holds, then
\begin{align}
\psi(E,x)\ &=\ \phi(E,x)\ +\ \sum_{l=1}^{N} \int_{\infty}^{x} {\phi(E,y)\ \phi_l(y)\ dy}\ \psi_l(x) \\
\big(H_N\ - E\big) \psi(E,x)\ &=\ \sum_{l=1}^{N} \int_{\infty}^{x} {\phi(E,y)\ \phi_l(y)\ dy}\ \big(H_N\ +\ \gamma_l^2\big)\psi_l(x)\ =\ 0  \label{}
\end{align} 
({\it i.e}) $\psi(E,x)$ is a solution of the Schr{\"{o}}dinger equation for the potential $U_N$ at energy $E$. It may then be shown that for positive energies $E=k^2$, the reflection and transmission coefficients of the two potentials $U_N$ and $U_0$ are related by
\begin{equation}
R_N(k)\ =\ R_0(k)\ \prod_{j=1}^N \frac{\gamma_j\ -\ i\ k}{\gamma_j\ +\ i\ k} \ ,\ \ \ \ T_n(k)\ =\ T_0(k)\ (-)^N\ \prod_{j=1}^N \frac{\gamma_j\ -\ i\ k}{\gamma_j\ +\ i\ k} \ .\label{}
\end{equation}

A relation satisfied by the Wronskians of the two sets of solutions $[\phi_k]$ and $[\psi_k]$ may be established by differentiating eq. (11), multiplying by $\psi_k$ and summing over $k$. These algebraic manipulations yield
\begin{equation}
\sum_{l=1}^N{\Big(\phi_l\ \frac{\partial \psi_l}{\partial x}\ -\ \psi_l\ \frac{\partial \phi_l}{\partial x}\Big)}\ =\ W^2  \ .\label{}
\end{equation}
The differentiation of eq. (12) gives
\begin{equation}
\sum_{l=1}^N {\Big(\phi_l\ \frac{\partial \psi_l}{\partial x}\ +\ \psi_l\ \frac{\partial \phi_l}{\partial x}\Big)}\ =\ -\frac{\partial W}{\partial x}\ =\ \frac{1}{2}\big(U_N - U_0\big)\ .   \label{}
\end{equation}
We can then establish that
\begin{align}
\sum_{l=1}^N {\phi_l\ \frac{\partial \psi_l}{\partial x}}\ &=\ +\frac{1}{2}\ \Big(W^2\ -\ \frac{\partial W}{\partial x}\Big) \ =\ +\frac{W^2}{2} \ +\ \frac{1}{4}\big(U_N-U_0\big)\ , \\
\sum_{l=1}^N {\psi_l\ \frac{\partial \phi_l}{\partial x}}\ &=\ -\frac{1}{2}\ \Big(W^2\ +\ \frac{\partial W}{\partial x}\Big)\ =\ -\frac{W^2}{2} \ +\ \frac{1}{4}\big(U_n-U_0)\ . \label{}
\end{align}
These results are valid for a general $U_0$ and are the generalizations of known results for the case $U_0=0$.

This procedure for adding $N$ bound states to a potential $U_0$ may be reversed so that starting from a potential with $N$ bound states in $U_N$ we can find a potential with no bound states in the form
\begin{align}
U_0\ &= \ U_N \ -\ 2\ \frac{\partial^2}{\partial x^2} {\ln{\det{B}}} \\
B_{kl} \ &= \delta_{kl} \ + \ {\int_{\infty}^x \ \psi_k(y) \ \psi_l(y) \ dy} \ =\ {\int_{-\infty}^x \ \psi_k(y) \ \psi_l(y) \ dy} \label{}
\end{align}
and the two sets of functions $\phi$ and $\psi$ are now related by
\begin{equation}
\sum_{l=1}^N{B_{kl}}\ \phi_l\ =\ \psi_k \ \ ,\ \ \phi_l\ =\ \sum_{k=1}^N{\big[B^{-1}\big]_{lk}\ \psi_k}  \label{}
\end{equation}
so that
\begin{equation}
-\sum_{l=1}^N \psi_l(x) \phi_l(x) = -\sum_{l=1}^N\sum_{k=1}^N\psi_l\big[B^{-1}\big]_{lk}\psi_k = -\sum_{l=1}^N\sum_{k=1}^N\big[B^{-1}\big]_{lk} \frac{\partial B_{lk}}{\partial x} = -\frac{\partial}{\partial x} {\ln{\det{B}}} .\label{}
\end{equation}

Comparison of the two sets of equations corresponding to the addition and the removal bound states shows that the matrix $[B]$ is the inverse of the matrix $[A]$ which means that in terms of the kernel defined by
\begin{equation}
K(x,y)\ =\ \sum_{k=1}^{N}{\psi_{k}(x)\ \phi_{k}(y)}  \label{}
\end{equation}
eqs. (11) and (30) may also be given in the form
\begin{align}
\psi_l(x) \ &=\ \sum_{k=1}^N \big[A^{-1}\big]_{lk}\phi_k\ =\ \sum_{k=1}^N B_{lk} \phi_k \ =\ \phi_l(x)\ +\ {\int_{\infty}^{x} \psi_l(y) \ K(y,x)\ dy} \\
\phi_l(x) \ &=\ \sum_{k=1}^N \big[B^{-1}\big]_{lk}\psi_k\ =\ \sum_{k=1}^N A_{lk} \psi_k \ =\ \psi_l(x)\ -\ {\int_{\infty}^{x}K(x,y)\  \phi_l(y)\ dy} \ .\label{}
\end{align} 
Eq. (19) may be expressed in terms of $K(x,x)$ in the form
\begin{equation}
U_N\ -\ U_0\ =\ -2\ \frac{\partial W}{\partial x} \ =\ 2\ \frac{\partial}{\partial x} K(x,x) \ .\label{}
\end{equation}

\section{Evolution in parameter space}

The potential $U_N$ constructed by adding bound states to $U_0$ may be viewed as the potential at time $t=0$. $U_0(x,0)$ and $U_N(x,0)$ may now be used to construct time dependent potentials $U_0(x,t)$ and $U_N(x,t)$ by letting both $U_0$ and $U_N$ satisfy one of the non-linear time evolution equations of the Lax hierarchy, given by eqs. (2) and (3), with the same value of $m$, and letting the eigenstates $[\phi_k]$ and $[\psi_k]$ evolve according to the appropriate set of eqs. (5) and (6). Under these circumstances the bound state eigenvalues remain unchanged while the bound state normalization constants and the reflection coefficients for positive energies for both potentials acquire a time dependence as discussed in section 1. However these changes will be the same for both potentials and the reflection coefficient and transmission coefficients for the two potentials will still be related in the manner indicated by eq. (23). Therefore, if we now eliminate the $N$ bound states of $U_N(x,t)$ by the procedure outlined in the last section the resulting potential will acquire a time dependence exactly of the form of $U_0(x,t)$. Only for the case of reflectionless category of potentials, for which the reflection coefficient vanishes for all $t$, $U_0(x,t)$ is zero for all $t$ and $U_0$ may be viewed as being time independent. This identification allows us to examine a different kind of parametric evolution in which $U_0(x)$ is a potential which in general has non-vanishing reflection coefficient, but remains independent of time, and the energies $[E_k]$ at which bound states are added are also independent of time.

Now suppose that a dependence of $[\phi_k]$ on a parameter $t$ can be introduced through a set of functions $[\alpha_k(t)]$ so that
\begin{equation}
\frac{\partial \phi_k}{\partial t} \ =\ \alpha_k(t)\ \phi_k\ \ ,\ \ \phi_k(x,t)\ =\ \exp\Big(\int_0^t\alpha_k({\bar t})\ d{\bar t}\Big)\ \phi_k(x,0) \label{}
\end{equation}
then $\phi_k(x,t)$ can still satisfy eq. (9) and $U_0$ may be chosen to be independent of $t$. The corresponding $t$ evolution of $[\psi_k(x,t)]$ may be examined by differentiating eq. (11) with respect to $t$ to get
\begin{equation}
\sum_{l=1}^N\Big(\frac{\partial A_{kl}}{\partial t}\ \psi_l\ +\ A_{kl}\ \frac{\partial \psi_l}{\partial t}\Big)\ =\ \alpha_k\ \phi_k \ .\label{}
\end{equation}
Since
\begin{equation}
\frac{\partial}{\partial t} A_{kl} \ =\ \big(\alpha_k \ +\ \alpha_l\big)\ \big(A_{kl} \ -\ \delta_{kl}\big) \label{}
\end{equation}
we can simplify eq. (37) using eq. (11) to the form
\begin{equation}
\sum_{l=1}^N A_{kl} \Big( \frac{\partial \psi_l}{\partial t} \ +\ \alpha_l \ \psi_l\Big)\ =\ 2 \alpha_k\ \psi_k \ .\label{}
\end{equation}
Hence the time evolution equation of $\psi_l$ is governed by
\begin{align}
\Big(\frac{\partial \psi_l}{\partial t} \ +\ \alpha_l  \psi_l\Big)\ &=\ 2\ \sum_{k=1}^N {\big[A^{-1}\big]_{lk} \alpha_k \psi_k}\ =\ 2\ \sum_{k=1}^N {B_{lk} \alpha_k \psi_k}\\
&= 2\int_{-\infty}^{x} {\psi_l(y,t) \sum_{k=1}^N \Big(\psi_k(y,t)\ \alpha_k(t)\ \psi_k(x,t)\Big)\ dy}\ . \label{}
\end{align}
Multiplication of eq. (39) by $\psi_k$, summation over $k$ and use of eqs. (11) and (36) then leads to the relation
\begin{equation}
\frac{\partial }{\partial t}\ \sum_{l=1}^N \psi_l \phi_l \ =\ -\frac{\partial W}{\partial t} \ = 2 \sum_{k=1}^N \alpha_k\ \psi_k^2 \label{}
\end{equation}
which when combined with eq. (19) leads to
\begin{equation}
\frac{\partial}{\partial t} \big(U_N \ -\ U_0\big) \ =\ \frac{\partial}{\partial t} U_N  \ = 4\frac{\partial}{\partial x}\ \sum_{k=1}^{N}{\alpha_k(t)\ \psi_k^2(x,t)}  \ .\label{}
\end{equation}
${\bullet}$ {\it Thus we have derived an evolution equation for $U(x,t)$ which is valid for arbitrary choices of values of the functions $[\alpha_k(t)]$ and depends on a sum over the bound state probabilities with weight factors $[\alpha_k]$.} 

Since $[\psi_k]$ are eigenfunctions normalized to unity, integration of eq. (42) between $-\infty$ and $\infty$ leads to the relation
\begin{equation}
Lt_{x\to -\infty}\ \frac{\partial}{\partial t} {\ln{\det{A}}}  \ =\ 2 \sum_{k=1}^N \alpha_k \label{}
\end{equation}
when the result from eq. (10) that as $x\to\infty$ $\det A\to 1$ is used.

Using the Schr{\"{o}}dinger equation satisfied by the eigenstates $\psi_k$ it can be shown that the probability density $\psi_k^2$ associated with the eigenstates of any potential satisfy
\begin{equation}
\Big(\frac{\partial^3}{\partial x^3} \ -\ 4 U_N \frac{\partial}{\partial x} \ -\ 2 \frac{\partial U_N}{\partial x} \Big) \ \psi_k^2 \ =\ 4 \gamma_k^2  \ \frac{\partial}{\partial x}\psi_k^2 \ ,\ k=1,2,..,N.\label{}
\end{equation}
The weighted sums
\begin{equation}
Q_j \ = \ -4\sum_{k=1}^N \beta_k \ \big(2\gamma_k\big)^{2j}\ \psi_k^{2}\ \ ,\   \ j=0,1,2,..\ \ ,\label{}
\end{equation}
where $[\beta_k]$ can take arbitrary values, therefore, satisfy 
\begin{equation}
\Big(\frac{\partial^3}{\partial x^3} \ -\ 4 U_N\  \frac{\partial}{\partial x} \ -\ 2 \frac{\partial U_N}{\partial x} \Big) \ Q_j \ =\ \frac{\partial Q_{j+1}}{\partial x} \ .\label{}
\end{equation}
If the time evolution functions in eq. (36) are chosen to be $[\alpha_k]= [\beta_k (2\gamma_k)^{2j}]$, then in terms of the members of the $[Q]$ hierarchy, eq. (43) becomes
\begin{equation}
\frac{\partial}{\partial t}U_N \ =\ -\frac{\partial Q_j}{\partial x} \ . \label{}
\end{equation}
Thus eqs. (36), (10), (11) (18-20) and  (46)-(48) generalize the equations for the Lax hierarchy. 

For the case $U_0=0$, the $N$-soliton potential $U_N =L_0$ where $L_0$ is the $m=0$ member of the Lax hierarchy defined by
\begin{equation}
L_m(x,t)\ =\ -2\ \sum_{k=1}^N \big(2\gamma_k\big)^{2m+1}\ \psi_k^2(x,t)\ \ \ ,\ \ \ m=0,1,2,...\ \ .\label{}
\end{equation}
Hence for the special case of the reflectionless potentials $U_N$ of the Lax hierarchy, constructed from $U_0=0$, one of the weighted sums in eq. (46), ({\it viz.}) $Q_0$, arising from the time independent choice of $[\beta_k(t)] =[\gamma_k]$, can be identified with the  $N$-soliton potential $U_N$. All other $[Q_j]$ can be expressed in terms of $U_N$ and its derivatives using eq. (47) and eq. (43) becomes
\begin{equation}
\frac{\partial U_N}{\partial t} \ =\ -\ \frac{\partial L_m}{\partial x} \label{}
\end{equation}
leading to the non-linear time evolution equation for the potential given by eq. (2). We have shown that for the case of $U_0=0$ the new parametric evolution discussed in this section becomes identical to the evolution equation of the Lax hierarchy given by eq. (2). This identity does not extend to potentials with non-vanishing reflection coefficients if we retain the choice of $[\beta_k]=[\gamma_k]$ and in general a new non-linear evolution equation arises which differs in essential respects from eqs. (2) and (3). 

We have studied how far it is possible to go in formulating parametric evolution equations, for potentials with non-vanishing reflection coefficients, which differ from the time evolution equations of the Lax hierarchy . There exists the possibility that in the general case, for a suitable choice of the functions $[\alpha_k(t)]$, which define the time development of the solutions in a time independent $U_0$, one of the weighted sums $[Q_j]$ can be related to a physically meaningful function. We examine such a possibility next.

\section{Green's function hierarchy and time evolution}

For the time independent choice 
\begin{equation}
\alpha_k(t)\ =\ \beta_k(t)\ =\ \frac{1}{4(\gamma_{k}^2+E)}\ ,\ \ E\ =\ -\gamma^2,\  \gamma \ne [\gamma_k], \label{}
\end{equation}
it is possible to relate $Q_0$ in eq. (46) to a Green's function by the following procedure. A Green's function $G(x,y;t,E)$ satisfying suitable boundary conditions may be defined as a solution of   
\begin{equation}
\Big(-\frac{\partial^2}{\partial x^2}\ +\ U(x,t) \ -\ E\Big) G(x,y;t,E)\ = \delta(x-y)  .\label{}
\end{equation}
It is well known that $G(x,y;t,E)$ may be expressed in terms of the complete set of eigenfunctions and eigenvalues of the Schr{\"{o}}dinger equation for the potential $U(x,t)$ (Morse and Feshbach 1953). For a general $U(x,t)$ the complete set of eigenfunctions include eigenstates for both discrete and continuous eigenvalues and $G$ may be represented by
\begin{equation}
G(x,y;t,E)\ =\ -\sum_{k}
\frac{\psi_{k}(x,t)\ \psi_{k}(y,t)}{\gamma_{k}^2+E}\ -\ \int{\frac{f(k)}{E-k^2} \psi^{\star}(k,x,t) \psi(k,y,t)\ dk}\   \label{}
\end{equation}
where $f(k)$ is the spectral density function for the potential $U$ for positive energy $E=k^2$. An alternative representation of $G$ can be given in terms of $\xi_{1}$ and $\xi_{2}$, the two linearly independent solutions of the Schr{\"{o}}dinger equation for $U$ at energy $E$,
\begin{equation}
\Big(-\frac{\partial^2}{\partial x^2} \ +U(x,t)\ -\ E\Big) \xi_{1,2}(x,t)\ =\ 0 \label{}
\end{equation}
chosen such that $\xi_{1}(x,t)\to 0$ as $x\to-\infty$ and $\xi_{2}(x,t))\to 0$ as $x\to\infty$. The Wronskian of the two solutions at the same energy E defined by 
\begin{equation}
\rho\ = \xi_{2}\ \frac{\partial \xi_{1}}{\partial x}\ -\ \frac{\partial \xi_{2}}{\partial x} \ \xi_{1}\  \label{}
\end{equation}
does not depend on $x$ as a consequence of the differential equation satisfied by $\xi_1$ and $\xi_2$. Then $G$ may be given in the form
\begin{equation}
G(x,y;t,E)\ =\  \ \frac{1}{\rho}\ \xi_{1}(x_{<},t)\ \xi_{2}(x_{>},t)  \label{}
\end{equation}
where $x_{<}\  (x_{>})$ is the smaller (larger) of $(x,y)$.
By following the same method as that used to prove eq.(45) it may be established that 
\begin{equation}
\Big(\frac{\partial^3}{\partial x^3} \ -\ 4 (U\ -\ E) \frac{\partial}{\partial x} \ -\ 2 \frac{\partial U}{\partial x} \Big) \ G(x,x;t,E) \ =\ 0 \ . \label{}
\end{equation}
If the potential $U$ is a confining potential with only a discrete spectrum then
\begin{equation}
Q_0\ =\ -\sum_{k} \frac{\psi_k^2(x,t)}{\gamma_k^2+E}\ =\ G(x,x;t,E)\ =\ \frac{\xi_{1}(x,t)\ \xi_2(x,t)}{\rho} \label{}
\end{equation}
and G(x,x;t,E) satisfies the sum rule
\begin{equation}
\int_{-\infty}^{\infty} G(x,x;t,E)\ dx\ =\ -\sum_{k}\frac{1}{\gamma_k^2\ +E} \ .\label{}
\end{equation}
Eq. (43) now becomes
\begin{equation}
\frac{\partial}{\partial t}U(x,t)\ =\ -\ \frac{\partial}{\partial x}G(x,x;t,E) \label{}
\end{equation}

Even if the potential is not a confining potential and has scattering states, the members of the hierarchy of functions defined by the discrete sums
\begin{equation}
G_j(x,y;t,E) \ = \ - \sum_{k=1}^N {\frac{1}{4^j}\ \frac{\psi_k(x,t)\psi_k(y,t)}{(\gamma_k^2+E)^{j+1}}} \ ,\ \ j=0,1,2,..\label{}
\end{equation}
satisfy
\begin{equation}
\Big(\frac{\partial^3}{\partial x^3} - 4\big(U_N-E\big)\frac{\partial}{\partial x} - 2 \frac{\partial U_N}{\partial x}\Big)G_{j+1}(x,x;t,E)\ = \ \frac{\partial}{\partial x}G_j(x,x;t,E)\ . \label{}
\end{equation}
Using the orthonormality of $[\psi_k]$ it is also possible to establish the integral expression
\begin{equation}
4 G_{j+1}(x,y;t,E)\ =\ \int_{-\infty}^{\infty} G_j(x,x_1;t,E) \ G_0(x_1,y;t,E)\ dx_1\ . \label{}
\end{equation}
Using eq. (43) it may be seen that the time dependent potential constructed from eqs. (36), (10) and (19) for the choice $\big[\alpha_k = (4\gamma_k^2+4E)^{-j-1}\big]$ with eigenstates given by eq. (11) and $G_j$ given by eq. (61) satisfies
\begin{equation}
\frac{\partial}{\partial t}U_N(x,t) \ =\ - \frac{\partial}{\partial x}G_j(x,x;t,E) \label{}
\end{equation}
$\bullet${\it This result is valid in general and this time evolution equation for $U$ in terms of the $[G]$ hierarchy exists for all potentials including the special case of $U_0=0$. In this new evolution equation $G_j$ plays the same role as $L_j$ does in the Lax hierarchy.} It is possible that other choices of $[\alpha_k]$ may exist for which the weighted sum $\sum\alpha_k\psi_k^2$ may represent a useful physical quantity.

In this paper we have shown that the methods used to establish the time evolution of the reflectionless potentials of the KdV equation can be extended to derive a parametric evolution equation for general potentials which is different from that of the Lax-KdV hierarchy. We have identified a generalized Lax hierarchy of functions and derived a new type of implicitly non-linear time evolution equation for general potentials which produces a (unitary?) time evolution of the eigenstates of the Schr{\"{o}}dinger equation for the potential.

\section{References}

\noindent[1] P.D.Lax, Comm. Pure. Appl. Math. {\bf 21}, 467 (1968).

\noindent[2] A.C.Scott, F.Y.E.Chu and D.W.Mclaughlin, Proc. I.E.E.E. {\bf 61}, 1443 (1973).

\noindent[3] M.Mulase, J. Diff. Geo. {\bf 19} 403 (1984).

\noindent[4] I.Kay and H.E.Moses, J. Appl. Phys. {\bf 27}, 1503 (1956).

\noindent[5] C.S.Gardner, J.M.Greene, M.D.Kruskal and R.M.Miura, Phys. Rev. Lett. {\bf 19}, 1095 (1967).

\noindent[6] H.B.Thacker, C.Quigg and J.L.Rosner, Phys. Rev. {\bf D18} 274, 287 (1978).

\noindent[7] P.J.Caudrey, R.K.Dodd and J.D.Gibbon, Proc.Roy.Soc.Lon. {\bf A351}, 407 (1976). 

\noindent[8] P.B.Abraham and H.E.Moses, Phys. Rev. {\bf A22}, 133 (1980).

\noindent[9] C.V.Sukumar, J. Phys. A: Math. Gen. {\bf 19} 2297 (1986).

\noindent[10] C.V.Sukumar, J. Phys. A: Math. Gen. {\bf 20} 2461 (1987).     

\noindent[11] D.Baye, J. Phys. A: Math. Gen. {\bf 20}, 5529 (1987).

\noindent[12] D.Baye and J.M.Sparenberg, Phys. Rev. Lett. {\bf 73}, 2789 (1994).

\noindent[13] P.Morse and H.Feshbach, 1953 {\it Methods of Theoretical Physics, Vol. 1}(New York: McGraw-Hill) 791-811.


\end{document}